\documentclass{ws-mpla}
\usepackage{graphicx, amssymb}
\newcommand{\D}{\displaystyle}

\makeatletter
 \newcommand{\figcaption}{\def\@captype{figure}\caption}
 \newcommand{\tabcaption}{\def\@captype{table}\caption}
\makeatother

\begin{document}

\markboth{L. S. Geng, H. Toki, and J. Meng} {A systematic study of
Zr and Sn isotopes in the Relativistic Mean Field theory}

%
%

\title{A systematic study of Zr and Sn isotopes in the Relativistic Mean Field theory}

\author{L. S. Geng}

\address{Research Center for Nuclear Physics (RCNP), Osaka
University,\\
Ibaraki, Osaka 567-0047, Japan\\
School of Physics, Peking University, Beijing 100871, P.R. China\\
lsgeng0@rcnp.osaka-u.ac.jp}

\author{H. Toki}

\address{Research Center for Nuclear Physics (RCNP), Osaka
University,\\
Ibaraki, Osaka 567-0047, Japan\\
toki@rcnp.osaka-u.ac.jp}

\author{J. Meng}
\address{
School of Physics, Peking University, Beijing 100871, P.R. China\\
mengj@pku.edu.cn } \maketitle


\begin{abstract}
The ground-state properties of Zr and Sn isotopes are studied
within the relativistic mean field theory. Zr and Sn isotopes have
received tremendous attention due to various reasons, including
the predicted giant halos in the neutron-rich Zr isotopes, the
unique feature of being robustly spherical in the region of
$^{100}$Sn $\sim$ $^{132}$Sn and the particular interest of Sn
isotopes to nuclear astrophysics. Furthermore, four (semi-) magic
neutron numbers, 40, 50, 82 and 126, make these two isotopic
chains particularly important to test the pairing correlations and
the deformations in a microscopic model. In the present work, we
carry out a systematic study of Zr and Sn isotopes from the proton
drip line to the neutron drip line with deformation effects,
pairing correlations and blocking effects for nuclei with odd
number of neutrons properly treated. A constrained calculation
with quadrupole deformations is performed to find the absolute
minimum for each nucleus on the deformation surface. All
ground-state properties, including the separation energies, the
odd-even staggerings, the nuclear radii, the deformations and the
single-particle spectra are analyzed and discussed in detail.

\end{abstract}


\section {Introduction}
One of the main aims of researches in nuclear physics is  to
describe the ground-state properties of all nuclei in the periodic
table with one method. Unfortunately, due to lack of understanding
in strong interaction and numerical difficulties in handling
nuclear many-body problems, so far all microscopic descriptions
are only possible on a phenomenological ground. The most
successful theories of this type are the conventional Hartree-Fock
method with effective density-dependent interactions
\cite{skyrme.56,gogny.80,floc.73} and its relativistic analog, the
relativistic mean field (RMF) theory
\cite{walecka.74,serot.86,reinhard.89}. The Hartree-Fock method is
based on the non-relativistic kinematics, in which the spin-orbit
interaction and the density dependent interaction are important
ingredients. While the RFM theory is based on the relativistic
kinematics, in which the nucleons, the mesons and their
interactions are the ingredients. Therefore, the spin appears
automatically and the interplay of the scalar and the vector
potentials leads naturally to a proper spin-orbit interaction and
appropriate shell structure. Such a property is very important for
the extrapolation of a phenomenological model to study exotic
nuclei, which have become accessible in recent years due to
developments in accelerator technology and detection techniques
\cite{tanihata.95}. The RMF theory has been very successful for
the description of many nuclear properties including binding
energies, nuclear radii and deformations from the proton drip line
to the neutron drip line \cite{hirata.91,ring.96}. However, the
number of nuclear properties investigated by the RMF model is
still considerably less than those done by its non-relativistic
counterpart.

To describe nuclear ground-state properties over a wide mass
region properly, we must take into account both deformation
effects and pairing correlations simultaneously. It has been known
for a long time that most nuclei except for a few with and near
magic numbers are deformed and most of them can be described by
axial deformations. In the present work, we adopt the expansion
method in the Harmonic-Oscillator basis to describe the
single-particle wave functions in the deformed mean field
potential \cite{gambhir.90}. In the original work of Gambhir. et
al. \cite{gambhir.90}, the pairing correlations are treated in the
BCS framework using a constant pairing interaction with a pairing
window. This method has been found to be incapable of describing
exotic nuclei where the continuum effects become important near
the drip lines. On the other hand, recently it was demonstrated
that the use of a zero-range $\delta$-force in the
particle-particle channel was able to take into account the
couplings to the continuum nicely in the non-relativistic mean
filed models \cite{sand.00,grasso.01}. This BCS framework with the
zero-range $\delta$-force has been introduced into the
relativistic mean filed models with the spherical shape
\cite{meng.96,meng.98,yadav.02,sand.03,yadav.04}. The zero-range
$\delta$-force acting on the paired nucleons in the BCS framework
is able to pick up the states (resonant states) having a large
overlap with the occupied states below the Fermi surface
\cite{sand.00,yadav.02,sand.03,yadav.04}.  We, therefore, have
introduced the zero-range $\delta$-force in the BCS framework into
the expansion method \cite{gambhir.90} in the deformed RMF model
and demonstrated its applicability in Ref. \cite{geng.031}. This
method has been applied to study the properties of light nuclei
\cite{geng.032}, particularly the proton-skins in the Ar isotopic
chain and the neutron-skins in the Na isotopic chain, and the
alpha-decay properties of the lately synthesized superheavy
elements 115 and 113 \cite{geng.034}. When applying this method to
the neutron-rich nuclei, due to their large spacial extension, we
do not expect the description of the sudden increase of neutron
radii in halo nuclei, which were found if we had solved the RMF
equations in coordinate space \cite{sand.03,geng.031,meng.982}.
However, except for the neutron radii of the halo nuclei, other
properties could be reproduced with satisfactory accuracy by this
method \cite{geng.031}. One way out of this deficiency for the
description of halo nuclei is to use the expansion method in the
Woods-Saxon basis \cite{ZMR03}, which unfortunately is only
possible for spherical nuclei at present.

The Zr and Sn isotopic chains have received lots of attention for
a long time. The Zr isotope has a semi-magic proton number 40, and
its neutron number can be (semi-) magic numbers 40, 50 and 82.
Most recently, several RMF calculations show that the so-called
giant halos may occur in the Zr isotopic chain where neutron
number goes beyond 82 \cite{sand.03,geng.031,meng.982}. The
evolution of deformation, the potential energy surfaces, the
microscopic structure of coexisting configurations, and shape
transitions in heavy Zr isotopes have been discussed extensively
since a long time ago (see Refs. \cite{naza.88,naza.92,naza.97}),
where sophisticated models have been employed and detailed
analysises have been made. In the present work, our discussions of
the evolution of the deformation in the Zr isotopic chain
(including that of the Sn isotopic chain) are restricted to the
mean field level. We would like to also compare our predictions
with those of two of the most successful non-relativistic mean
field models, the finite-range droplet model (FRDM)
\cite{poller.95} and the Hartree Fock + BCS method (HF+BCS)
\cite{goriely.01}. We notice that the effect of the proton-neutron
pairing for $^{80}$Zr is also revisited lately \cite{goodman.99}.

The Sn isotope has long been of special interest for nuclear
physicists not only because it has a magic proton number 50, but
also due to its extremely long isotopic chain. Experimentally both
the double magic nuclei, the neutron-deficient $^{100}$Sn and the
neutron-rich $^{132}$Sn, have been produced. We expect that its
isotopes will even extend to the next neutron magic number,
$N=126$, which will be one of the objects of the present study.
Most Sn isotopes have been found to be spherical both
experimentally and theoretically, particularly those nuclei with
$N=50\sim82$, but whether or not such a trend can continue to the
next neutron-magic number, $N=126$, remains to be an interesting
and yet open question. Recently, Samanta et al. suggested that the
Pb isotope ($Z=82$) loses its magicity around $N=106$ based on the
extended Bethe-Weizs\"{a}cker mass formula \cite{samanta.03}. This
motivates us to study also the magicity of $Z=50$ by the RMF
method. The RMF model is particularly suitable for such a purpose
because of its natural description of spin-orbit interaction. Just
like Zr isotopes, Sn isotopes have also been investigated quite a
lot both in relativistic \cite{meng.99,meng.992,lala.98} and
non-relativistic mean field \cite{sakakihara.01} models . However,
most of them are limited to either spherical assumptions, or
even-even nuclei, or just part of the isotopic chain. The same
limitations are also true for Zr isotopes.

In the present work, we apply the recently formulated deformed
RMF+BCS method with a zero-range $\delta$-force in the pairing
channel \cite{geng.031,geng.032,geng.034} to the analysis of the
ground-state properties of Zr and Sn isotopes from the proton drip
line to the neutron drip line. In particular, we would like to
study the neutron-rich nuclei in both isotopic chains and the
magicity of $Z=50$ in the neutron-rich region. We use the TMA
\cite{sugahara.94} and NL3 \cite{lalazissis.97} parameter sets for
the RMF Lagrangian density, which are two of the most successful
parameter sets in the relativistic mean field model. In Sec. 2, we
briefly present the RMF model with deformation effects and pairing
correlations. The numerical details and discussions will be
presented in Sec. 3 and a summary of the present study is given in
Sec. 4.
\section{Relativistic mean field model with deformation effects and
pairing correlations}
 We briefly present here the formulation of the RMF
model with deformation effects and pairing correlations.  We
employ the model Lagrangian density with nonlinear terms for both
$\sigma$ and $\omega$ mesons, as described in detail in Ref.
\cite{geng.031,sugahara.94}, which is given by
\begin{equation}
\begin{array}{lll}
\mathcal{L} &=& \bar \psi (i\gamma^\mu\partial_\mu -M) \psi +
\,\frac{\D 1}{\D 2}\partial_\mu\sigma\partial^\mu\sigma-\frac{\D
1}{\D 2}m_{\sigma}^{2} \sigma^{2}- \frac{\D 1}{ \D
3}g_{2}\sigma^{3}-\frac{\D 1}{\D
4}g_{3}\sigma^{4}-g_{\sigma}\bar\psi
\sigma \psi\\
&&-\frac{\D 1}{\D 4}\Omega_{\mu\nu}\Omega^{\mu\nu}+\frac{\D 1}{\D
2}m_\omega^2\omega_\mu\omega^\mu +\frac{\D 1}{\D
4}g_4(\omega_\mu\omega^\mu)^2-g_{\omega}\bar\psi
\gamma^\mu \psi\omega_\mu\\
 && -\frac{\D 1}{\D 4}{R^a}_{\mu\nu}{R^a}^{\mu\nu} +
 \frac{\D 1}{\D 2}m_{\rho}^{2}
 \rho^a_{\mu}\rho^{a\mu}
     -g_{\rho}\bar\psi\gamma_\mu\tau^a \psi\rho^{\mu a} \\
      && -\frac{\D 1}{\D 4}F_{\mu\nu}F^{\mu\nu} -e \bar\psi
      \gamma_\mu\frac{\D 1-\tau_3}{\D 2}A^\mu
      \psi,\\
\end{array}
\end{equation}
where the field tensors of the vector mesons and of the
electromagnetic field take the following forms:
\begin{equation}
\left\{
\begin {array}{rl}
\Omega_{\mu\nu} =&
\partial_{\mu}\omega_{\nu}-\partial_{\nu}\omega_{\mu}\\
 R^a_{\mu\nu} =& \partial_{\mu}
                  \rho^a_{\nu}
                  -\partial_{\nu}
                  \rho^a_{\mu}-2g_\rho\epsilon^{abc}\rho^b_\mu\rho^c_\nu\\
 F_{\mu\nu} =& \partial_{\mu}A_{\nu}-\partial_{\nu}
A_{\mu}
\end{array}\right.
\end{equation}
and other symbols have their usual meanings. Based on the
single-particle spectra calculated by the RMF method, we perform a
state-dependent BCS calculation \cite{geng.031,lane.64,ring.80}.
The gap equation has a standard form for all the single particle
states. i.e.
\begin{equation}\label{eq:bcs}
\Delta_k=-\frac{\D 1}{\D 2}\sum_{k'>0}\frac{\D \bar{V}_{kk'}
\Delta_{k'}}{\D\sqrt{(\varepsilon_{k'}-\lambda)^2+\Delta_{k'}^2}},
\end{equation}
where $\varepsilon_{k'}$ is the single particle energy and
$\lambda$ is the Fermi energy. The particle number condition is
given by $2\sum\limits_{k>0} v_k^2=N$. As have been done in the
recent works \cite{geng.031,geng.032,geng.034}, a zero-range
$\delta$-force $V=-V_0\delta({\bf r_1}-{\bf r_2})$ is used in the
particle-particle channel in order to take into account the
continuum effects properly. For nuclei with odd number of
nucleons, a simple blocking method without breaking the
time-reversal symmetry is adopted. The ground state of an odd
system is described by the wave function,
\begin{equation}
\alpha^\dagger_{k_1}|BCS>=\alpha^\dagger_{k_1}\prod_{k\ne
k_1}(u_k+v_k\alpha^\dagger_k\alpha^\dagger_{\bar{k}})|vac>.
\end{equation}
Here, $|vac>$ denotes the vacuum state.  The unpaired particle
sits in the level $k_1$ and blocks this level. The Pauli principle
prevents this level from participating in the scattering process
of nucleons caused by the pairing correlations. As described in
Ref. \cite{geng.032,ring.80}, the gap equation with one level
$k_1$``blocked" has the form:
\begin{equation}
\Delta_k=-\frac{\D 1}{\D 2}\sum_{k'\ne k_1>0}\frac{\D
\bar{V}_{kk'}
\Delta_{k'}}{\D\sqrt{(\varepsilon_{k'}-\lambda)^2+\Delta_{k'}^2}},
\end{equation}
with the corresponding chemical potential determined by $
N=1+2\sum\limits_{k\ne k_1>0}v^2_k$. This blocking procedure is
performed at each step of the self-consistent iteration and more
numerical details will be discussed below.

\section{Numerical details and discussions}
\subsection{The convergence study} For
heavy nuclei more Harmonic-Oscillator shells  in the expansion of
fermion and boson fields have to be used in order to achieve
reliable convergence. However, computation time increases
dramatically with the increase of the number of shells. The common
procedure to tackle this problem is to limit the number of shells
by studying the convergence behavior in the region of interest.
For this reason, we investigate in detail three nuclei $^{132}$Sn,
$^{176}$Sn and $^{130}$Zr, which are chosen to represent spherical
neutron-rich nuclei, spherical drip-line nuclei and deformed halo
nuclei. For the case of $^{132}$Sn, with the basis parameter
$\beta_0=0.0$, it is found that for $N=N_f=N_b\ge14$, where $N_f$
and $N_b$ are the numbers of shells used in the expansion of
fermion and boson fields, the desirable convergence is obtained
properly. The increase of $N$ from 14 to 20 changes the total
binding energy $B$ and the binding energy per particle $B/A$ by
about $0.2\%$, while the rms charge radii $R_c$ is changed by
$0.1\%$ and the rms neutron radii $R_n$ is changed by $0.3\%$. For
$^{132}$Sn, even 12 shells can obtain the satisfactory convergence
as shown in Table I. Similar conclusion can be made for $^{176}$Sn
(Table II) except that now a minimum of 14 shells is needed.
Things are a little bit different for $^{130}$Zr (Table III),
which has been predicted to be a halo nucleus \cite{meng.982}.
With $N$ increasing from 14 to 20, $B$ and $B/A$ change by about
$0.6\%$, while $R_c$ is changed by $0.1\%$, and $R_n$ is changed
by $0.7\%$. In our calculations the ground state of $^{130}$Zr has
a deformation of $\beta_{2m}=-0.203$. Setting $\beta_0$ equal to
$-0.209$, we find that $\beta_{2m}$ is changed by about $3\%$ when
$N$ increases from 14 to 20.

In conclusion, we find that 14 shells are enough for a correct
description of $B$, $B/A$, $R_c$ and $\beta_2$. While for $R_n$,
to describe the halo nuclei more reliably, shells more than 14 are
needed. Of course, as is well known (see Ref. \cite{ZMR03} and
references therein), the rms neutron radii of neutron halo nuclei
can not be fully reproduced within the Harmonic-Oscillator basis
even with a large number of basis. Keeping all these in mind, we
decide to use 20 shells ($N_f=N_b=20$) for the calculations with
the TMA parameter set, while for the calculations with the NL3
parameter set 14 shells ($N_f=N_b=14$) are used to save
computation time.

\subsection{The blocking method and the pairing strength}
As we have stated in Sec. 2, for nuclei with odd number of
neutrons a blocking method without breaking the time-reversal
symmetry is adopted in the self-consistent calculation. At each
step, the state to be blocked is determined by counting the
single-particle spectra from the bottom. Then with this state
blocked, the BCS calculation is performed. This procedure is
repeated until the required convergence is obtained. The same
procedure is also adopted in the constrained calculations. For
each nucleus, first the constrained calculation \cite{flocard.73}
is done to obtain all possible ground-state configurations; second
we perform the non-constrained calculation using the quadrupole
deformation parameter of the deepest minimum of the energy curve
of each nucleus as the deformation parameter for the
Harmonic-Oscillator basis. In the case of several similar minima
from the constrained calculation, the non-constrained calculation
is repeated around each minimum to obtain the configuration with
the lowest energy as the final result.

We note that the most appropriate way to perform blocking
calculations is by blocking all orbits accessible to the odd
nuclei at each deformation, but it is incredibly time consuming.
Therefore, our method should be seen as an approximation to that
kind of full calculation and the agreement should be checked at
certain points. Here we take $^{151}$Sn in our RMF+BCS
calculations with the TMA parameter set as an example to show the
validity of this method. First, the constrained calculation finds
that the deepest minimum is at $\beta_2\approx -0.263$. Using this
as the basis deformation, i.e. $\beta_0=-0.263$, we perform the
non-constrained calculation. The final result is $B=1142.791$ with
$\beta_2=-0.266$, where $B$ is the total binding energy. Looking
at the single-particle spectra, we find that the blocked state is
$1/2^-[521]$ with the single-particle energy $E=-2.239$. Two other
possible states accessible to the odd neutron is $3/2^-[521]$ with
$E=-2.374$ and $11/2^+[615]$ with $E=-2.156$. If one of these two
states were blocked instead of $1/2^-[521]$, the results would be
$B=1142.728$ with $\beta_2=-0.267$ and $B=1142.659$ with
$\beta_2=-0.267$. These are all close to each other and our method
chooses the correct configuration with the lowest energy. This
case shows that the procedure we adopted works very well. However,
there are cases where the state chosen by this procedure does not
lead to the ``absolute minimum''. Nevertheless, such wrong
selections of the blocking states generally lead to a difference
less than 0.2 MeV for the total binding energy, which is actually
beyond the accuracy of our calculations and therefore can be
safely ignored.

The pairing strength $V_0=341.1$ Mev fm$^3$  is used for Zr
isotopes and $V_0=310.0$ MeV fm$^3$ for Sn isotopes, which are
obtained by fitting the experimental odd-even staggerings (OESs)
of Zr and Sn isotopes with a cut-off $E_{max}-\lambda\le8.0$ MeV.
We use the same pairing strength for both protons and neutrons.
For comparison, calculations have been done with the NL3 parameter
set also, but only for even-even nuclei. The pairing strength is
kept the same for calculations with both the TMA parameter set and
the NL3 parameter set.

\subsection{Neutron separation energies and binding energies}
The double and single neutron separation energies
\begin{equation}
S_{2n}(Z,N)=B(Z,N)-B(Z,N-2),
\end{equation}
\begin{equation}
S_{n}(Z,N)=B(Z,N)-B(Z,N-1),
\end{equation}
are displayed in Figs. 1 and 2 for Zr isotopes and Sn isotopes,
respectively, from the proton drip line to the neutron drip line.
The results obtained from the deformed RMF+BCS calculations with
the parameter set TMA are compared with available experimental
data \cite{audi.95}. In the upper panels of Figs. 1 and 2, the
results obtained from the deformed RMF+BCS calculations with the
NL3 parameter set are also shown for comparison. It is clearly
seen that both experimental separation energies and OESs are
reproduced quite well by the theoretical calculations. The RMF+BCS
calculations with different parameter sets TMA and NL3 also agree
with each other very well.

For Zr isotopes, despite of the fact that the variation of the
deformation is quite complicated for the whole isotopic chain (see
Figs. 4 and 6), the agreement between theory and experiment is
very good. The experimental feature that the OESs are larger for
nuclei with $N=40\sim50$ than for nuclei with $N=51\sim68$ is also
reproduced quite well. Both the double and single neutron
separation energies are quite small for nuclei with $N>82$. The
so-called giant halos are predicted to exist in this region by
both the RCHB \cite{meng.982} and the resonant RMF+BCS
\cite{sand.03} calculations with spherical assumptions. Our
previous deformed RMF+BCS calculations \cite{geng.031} reached the
same conclusion and confirmed that the expansion method in the
Harmonic-Oscillator basis can provide very accurate results for
nuclear binding energies even in the neutron-rich region. All
these calculations\cite{sand.03,geng.031,meng.982}, however, are
limited to only even-even nuclei. From the lower panel of Fig. 1,
we notice that the OESs of Zr isotopes with $N>82$ are quite small
and $S_n$ of those nuclei with odd mass numbers are almost zero or
negative. This is easy to understand because the pairing
correlations can increase the binding energies of even-even nuclei
more and therefore contribute to the formation of halo nuclei. We
also conclude that generally we will see one-neutron halo nucleus
first and then only two-neutron halo nucleus is stable due to the
pairing correlations. The same conclusion also holds for Sn
isotopes.

For Sn isotopes, we notice that the agreement between theory and
experiment is very good in the region, $^{115}$Sn $\sim$
$^{132}$Sn. The two-neutron drip-line nucleus is predicted to be
$^{176}$Sn by our calculations with both the TMA and the NL3
parameter sets, which agree well with the calculations with other
parameter sets in the RMF framework \cite{long.03}. This seems to
be a unique feature of the RMF calculations while the HF
calculations usually shift the drip-line nucleus toward the light
mass region (see Fig. 4 of Ref. \cite{naza.99}). Different isospin
dependence and spin-orbit description could be the reason why this
discrepancy happens. However, since we are talking about exotic
nuclei far from the stability region, this remains to be an open
question anyway. For nuclei with $N\le64$, the theoretical OES
drops a little bit suddenly. The reason behind this is a
relatively large spin-orbit splitting between $2d_{5/2}$ state and
$2d_{3/2}$ state, around 1.8 MeV in our present calculations (see
Figs. 7 and 8), which makes pairing energy in nuclei with $N\le64$
smaller than in nuclei with $N=64\sim82$. We note that the smaller
OES in the region of $^{146}$Sn $\sim$ $^{160}$Sn is quite
complicated and may be related with two different phenomena. One
is that nuclei in this region are deformed (see Fig. 5). The other
one is that as we shall see from Fig. 8, the $3p_{1/2}$,
$2f_{5/2}$ and $3p_{3/2}$ states, which become occupied in this
region, are quite close to each other. In both Zr and Sn isotopic
chains, we should note that for nuclei with $N\approx Z$
additional residual interactions, such as the proton-neutron
pairing, will contribute and increase the OES somewhat, which are
not included in the present model. Unlike Zr isotopes, we do not
see a clear indication of possible two-neutron halo in the Sn
isotopic chain, which is consistent with all previous
calculations. The double neutron separation energy goes directly
to about $-2.5$ MeV for $^{178}$Sn from about 1.4 MeV for
$^{176}$Sn. However, from the lower panel of Fig.2, we can see
that $S_n$ stays below 1.0 MeV from $^{157}$Sn till the end of
this isotopic chain, which implies the possible existence of
one-neutron halo.

The average binding energy per particle is plotted as a function
of mass number A in Fig. 3 for Zr and Sn isotopes. It can be
clearly seen that the theoretical predictions reproduce the
experimental data accurately including the OESs in Sn isotopes.
For Zr isotopes, the small deviation between theory and experiment
for nuclei with $N=52\sim58$ is probably related with the particle
number non-conservation feature of the present BCS method, which
will be discussed more below. For Sn isotopes around $N\approx Z$,
our calculated results are somewhat larger than the experimental
data.

\subsection{Deformation properties}
Deformation is quite an important property for nuclei. It can
increase the nuclear binding energy so that a deformed shape is
more favored than the spherical shape for certain nuclei. The
deformation effect is also important for the description of
odd-even staggering (OES). It has been argued that in light- or
medium-mass nuclei deformation provides the same amount of OES as
the pairing interaction \cite{satula.98} and/or that the OES in
light nuclei is a mere reflection of the deformed mean field
\cite{hakk.97}. Although it is believed that the OES in heavy
nuclei is mainly due to the pairing correlations, the deformation
effect should never be ignored. We have also shown in Ref.
\cite{geng.031} that the deformation effect in even-even Zr
isotopes is quite important to reduce the discrepancy between the
theoretical and the experimental binding energies. $Z=50$ has long
been seen as a special magic number in the sense that all Sn
isotopes are believed to be spherical. Whether or not Sn isotopes
are deformed also is of special interest to the synthesis of
nuclei in the astrophysical r-process.

In Figs. 4 and 5, we plot the mass quadrupole deformation
parameters, $\beta_{2}$, for Zr and Sn isotopes. The results of
our calculations with the TMA and NL3 parameter sets are compared
with those of the finite-range droplet model (FRDM)
\cite{poller.95} and the Hatree Fock + BCS (HF+BCS)
\cite{goriely.01} method. The experimental data are extracted from
the B(E2) values given in Ref. \cite{raman.01}.

From Fig. 4, it is seen that the variation of the deformation is
quite complicated for Zr isotopes. The agreement between different
theoretical calculations, RMF+BCS/TMA, RMF+BCS/NL3, FRDM and
HF+BCS, and experimental data is quite good in general. Nuclei
with $N\le80$ are very deformed in all the theoretical
calculations. The same is true for nuclei with $58\le N\le 70$.
Nuclei with $N>88$ are moderately deformed with
$\beta_2\approx0.3$ by FRDM and $\beta_2\approx-0.3$ by RMF+BCS
with both the TMA and NL3 parameter sets, while HF+BCS seems to
prefer spherical shapes for this region. There are two regions
where the theoretical calculations do not agree so well with each
other. The first one is around $42\le N\le 54$, where the results
of FRDM and RMF+BCS/NL3 stay almost spherical while HF+BCS and
RMF+BCS/TMA show signs of oscillation between oblate and prolate
shapes. The second region is about $70<N<80$, where both
RMF+BCS/TMA and FRDM predict oblate shapes while RMF+BCS/NL3 and
HF+BCS prefer spherical configurations. These discrepancies
between different theoretical calculations indicate that the
nuclei in these regions are ``soft'' in the sense of deformation
(see also Fig. 6).

However, Sn isotopes are quite different from Zr isotopes. It
remains robustly spherical till $N$ reaches 88 in all the
theoretical calculations. Then the non-relativistic calculations
FRDM and HF+BCS show a region of moderately deformed with
$\beta_2\approx 0.3$ around $90<N<110$. The relativistic
calculations show a transition from prolate shapes ($90<N<100$) to
oblate shapes ($100<N<110$) while the amplitudes of deformation
are similar to their non-relativistic counterparts. For nuclei
with $N$ approaching the magic number 126, the RMF+BCS
calculations and the HF+BCS calculations predict spherical shapes
while the FRDM calculations prefer prolate configurations. It is
quite a surprise that all the theoretical works, either
non-relativistic or relativistic, predict a deformed region around
$90<N<110$ for Sn isotopes , which makes our predictions well
grounded. Although it may be trivial, another well-known feature
shown in our calculations is that the even-odd nucleus tends to be
deformed despite that its neighboring even-even nuclei are
spherical. Of course at most times, the deformation is quite small
and we can say that it is essentially spherical. Most even-odd Sn
isotopes with $N<82$ belong to this group. However, in Zr
isotopes, some even-odd nuclei differer from their even-even
neighbors by an absolute $\beta_{2}$ of 0.1. $^{87}$Zr, $^{89}$Zr,
$^{117}$Zr and $^{119}$Zr are such nuclei.

Usually, we can find two minima on the total energy surface of an
axially deformed nucleus, one prolate and the other oblate, or
spherical. On the theoretical side, it is easy to tell which one
is the ground-state configuration if the difference between the
total energies of different minima is relatively large. On the
other hand, nuclei in transition region often have two minima with
similar total energies, where it is difficult to determine the
ground state without ambiguity. $^{82}$Zr, $^{86}$Zr, $^{93}$Zr
and $^{116}$Zr are such nuclei in the Zr isotopic chain. We
display their total energy curves together with those of 12 other
Zr isotopes as functions of $\beta_{2}$ in Fig. 6. It provides us
a vivid picture how deformation evolves in Zr isotopes. Nuclei
with $A<82$ are strongly prolate with $\beta_{2}\approx0.5$ . At
$^{82}$Zr, the oblate configuration becomes as stable as the
prolate configuration. Nuclei with $82<A<86$ are moderately oblate
with $\beta_{2}\approx-0.2$. At $^{86}$Zr, the total energy curve
becomes quite flat at the bottom and the spherical configuration
is a little bit more stable. With increasing $N$, even-even nuclei
stay almost spherical while even-odd nuclei are somewhat deformed.
At $^{93}$Zr, the prolate configuration $\beta_{2}\approx0.2$ is
more stable than the oblate configuration $\beta_{2}\approx-0.1$.
Nuclei with increasing $N$ become more prolate and $\beta_2$
reaches 0.4 at $^{109}$Zr. Then $^{110}$Zr becomes oblate and
$\beta_2$ begins to approach zero with increasing $N$. At
$^{116}$Zr, the spherical configuration is reached and maintained
for even-even nuclei till $^{124}$Zr. Then the nucleus with more
neutrons become more prolate till $^{129}$Zr and from $^{130}$Zr
it stays moderately oblate, $\beta_{2}\approx-0.2$, till the end
of this isotopic chain. It will be very interesting to study
shape-coexistences in the whole Zr isotopic chain within the
relativistic mean field model just as what have been done in the
non-relativistic models (see Refs. \cite{naza.92,naza.97}) and
compare with experimental data, if available. (Here it should be
noted that the different minima on the potential energy surfaces
may be linked by a valley in the gamma angle, so that a secondary
minimum might imply a gamma-soft structure or even indicate a
triaxial structure, which of course is not supposed to be seen in
the present work due to the assumption of axial symmetry.)

We would like to make a few more comments about the discrepancy
between theory and experiment in the region of $N=52\sim58$ in the
Zr isotopic chain. Experimentally \cite{raman.01}, the nuclei
$^{92}$Zr, $^{94}$Zr, $^{96}$Zr and $^{98}$Zr are almost spherical
and nuclei with $N\ge60$ suddenly become very deformed. While in
our calculations (including those of FRDM and HF+BCS models), the
nucleus becomes deformed gradually from $N=52$ with $N$ and agrees
with experiment after $N=60$. The reason could be that the current
BCS method mixes the wave functions with different particle number
and therefore leads to finite deformation for the nuclei in the
region of $N=52\sim58$. A calculation with an improved BCS method
that conserves the particle number could clarify this discrepancy
and this work is in progress now.

\subsection{Single-particle spectra of Sn isotopes}
 It has long been known that
the nuclear deformation is caused by the interplay between protons
and neutrons ( see Ref. \cite{naza.92} and references therein). In
this sense, the reliability of the deformation developments
predicted by our calculations are more or less determined by to
what extent the relativistic mean field theory with the particular
parameter sets TMA and NL3 can reproduce the experimental
single-particle spectra, particularly the proton gap at $Z=50$ and
the neutron gap at $N=82$. In Fig. 7, the single-particle energies
of $^{132}$Sn calculated with the parameter sets TMA and NL3 are
compared with the experimental single-particle energies
\cite{long.03,rutz.98} for protons (left panel) and neutrons
(right panel). It is very clear that the calculations with both
the TMA parameter set and the NL3 parameter set agree with each
other very well and are also in reasonable agreement with the
experimental single-particle spectra. The proton gap at $Z=50$ is
reproduced quite well by NL3, while TMA predicts a gap of 0.5 MeV
smaller than experiment. Meanwhile, the calculated neutron gap at
$N=82$ is somewhat larger than experiment for both parameter sets.
We should note that such a comparison is not very well defined and
a detailed analysis of the single-particle spectra in the
relativistic mean field theory can be found in Ref.
\cite{rutz.98}. Anyway, the good agreement with the experimental
data justifies us to extend the relativistic mean field model to
the study of exotic nuclei.

In our above discussions, some small discrepancies in the
single-neutron separation energies for Sn isotopes with $N<64$
have been attributed to their single-particle spectra. The
abnormalities in the region of $^{146}$Sn $\sim$ $^{160}$Sn are
also related to the corresponding shell structures. To see all
these more clearly, we display the neutron single-particle spectra
of five nuclei in Sn isotopes, i.e. $^{100}$Sn, $^{114}$Sn,
$^{132}$Sn, $^{158}$Sn and $^{176}$Sn in Fig. 8. These isotopes
are so chosen to represent the whole Sn isotopes from the proton
drip line to the neutron drip line and also include two typical
nuclei in the region where the predicted OES changes abruptly. All
the five nuclei are spherical in our calculations. It is quite
clear that the $1h_{11/2}$, $3s_{1/2}$ and $2d_{3/2}$ states are
quite close to each other, which explains the somewhat large
difference in the OES between nuclei with $N>64$ and $N<64$,
together with the relatively large gap between the $2d_{3/2}$
state and the $2d_{5/2}$ state. The small gap between the
$3p_{1/2}$, $2f_{5/2}$ and $3p_{3/2}$ states is possibly
responsible for both the deformation phenomena and abnormalities
in the OES in the region of $^{146}$Sn $\sim$ $^{160}$Sn.

There are also some other important features shown in Fig. 8.
First, the neutron level density increases with $N$: the
single-particle energy of the $1s_{1/2}$ state increases about
10.0 MeV from $-62.0$ MeV in $^{100}$Sn to $-52.0$ MeV in
$^{176}$Sn; meanwhile, the single-particle energy of the
$1g_{9/2}$ state decreases about 2.0 MeV. Second, it is quite
clear that the spin-orbit splitting also decreases with increasing
$N$. Some obvious spin-orbit partners, which exhibit this feature
well, are $(1p_{1/2},1p_{3/2})$, $(1d_{3/2},1d_{5/2})$ ,
$(1f_{5/2},1f_{7/2})$ and $(1g_{7/2},1g_{9/2})$. The decrease of
spin-orbit splitting with increasing $N$ in the relativistic mean
field theory has been discussed in several previous works (see
Ref. \cite{meng.992}), where it has been attributed to the
diffuseness of the potential.
\subsection{Nuclear radii and neutron skin}

From the proton (neutron) density distributions, $\rho_{p(n)}$, we
can easily deduce the corresponding proton (neutron) radius,
\begin{equation}
R_{p(n)}=\langle r^2_{p(n)}\rangle^{1/2}= \left[\frac{\int
\rho_{p(n)} r^2d{\bf r}}{\int \rho_{p(n)} d{\bf r}}\right]^{1/2}.
\end{equation}
The charge radius, $R_c$, is related to the proton radius, $R_p$,
after considering the finite size of the proton, by
\begin{equation}
R^2_c=R^2_p+0.64 \quad\quad \mbox{[fm]}.
\end{equation}
We display in Figs. 9 and 10 the charge, proton and neutron rms
radii for Zr and Sn isotopes. The calculated charge radii are
compared with the experimental data from Ref. \cite{veris.87}.
Excellent agreement between theory and experiment can be clearly
seen. We can also see the effect of deformation clearly from Figs.
9 and 10. In the upper panel of Fig. 9, $R_c$ of Zr isotopes drops
dramatically for $^{82}$Zr and $^{110}$Zr, which is due to the
deformation effect. If we take a look at Fig. 4, it becomes much
clearer: $\beta_{2p}$ changes from 0.535 for $^{81}$Zr to $-0.218$
for $^{82}$Zr; correspondingly, $\beta_{2p}$ changes from 0.418
for $^{109}$Zr to $-0.216$ for $^{110}$Zr. There are also some
abnormalities in Sn isotopes as we can see in the upper panel of
Fig. 10, which have the same origins as Zr isotopes. $\beta_{2p}$
jumps from 0.085 for $^{144}$Sn to 0.220 for $^{145}$Sn and it
changes from $-0.244$ for $^{157}$Sn to 0.0 for $^{158}$Sn.

In the lower panels of Figs. 9 and 10, we plot $R_n$ and $R_p$
together in order to see the differences between them easily. We
also display a solid line with a $N^{1/3}$ dependence in both
panels for $R_n$. It is quite clear that the neutron-rich nuclei
in Zr isotopes deviate from the $N^{1/3}$ line quite a lot, which
implies that the giant halos exist even though deformation
develops after $N=82$. While for Sn isotopes our calculated
results agree well with the $N^{1/3}$ line. The difference between
$R_n$ and $R_p$, i.e. neutron skin, reaches 0.9 fm for $^{140}$Zr
and 0.8 fm for $^{176}$Sn, which can be viewed as another
indication of possible neutron halo in Zr isotopes. Of course, we
should note that as we have found in our previous work
\cite{geng.031}, the sudden increase of the neutron radii in
$^{124}$Zr $\sim$ $^{140}$Zr, predicted by the RCHB
\cite{meng.982} and the resonant RMF+BCS \cite{sand.03}
calculations, can not be fully reproduced by the expansion method
in the Harmonic-Oscillator basis. Nevertheless, this should not
prevent us from gaining some useful insights into possible neutron
halo feature, as we can see in Zr isotopes where a sudden increase
in neutron radii from $^{123}$Zr is quite obvious.

\section{Conclusions}
It is our strong desire to develop a model valid for all nuclei,
including unstable ones from the proton drip line to the neutron
drip line. The relativistic mean field theory, as a relativistic
method which incorporates spin-orbit interaction naturally,
hopefully can be the desirable model with only a few parameters
fitted by the saturation properties of nuclear matter and the
ground-state properties of a few well-known spherical nuclei.
However, to describe nuclei over a wide mass region, deformation
effects and pairing correlations must be taken into account
simultaneously and in a proper way. The present method we adopted
can fulfill all these requirements nicely and cost reasonable time
to carry out large-scale calculations covering all nuclei from the
proton drip line to the neutron drip line, which are necessary to
test a phenomenological method and provide reliable information
for both experimental physicists and other related subjects.

In the present work, we have studied two of the most interesting
isotopic chains in the periodic table, Zr and Sn, from the proton
drip line to the neutron drip line. This is the first of such kind
of calculations carried out in the relativistic mean field model
for the whole Zr and Sn isotopic chains including deformation
effects and proper pairing correlations. Our calculations
reproduce the available experimental data very well including the
nuclear binding energies, the separation energies, the odd-even
staggerings (OESs), and the nuclear radii. In particular, nuclear
deformations are analyzed in detail for both Zr and Sn isotopes in
the mean field level. Remarkable agreements with the very
successful non-relativistic theoretical models, the FRDM model and
the HF+BCS method, are observed. Surprisingly, we find that Sn
isotopes are deformed in the region of $N=89\sim 107$. The fact
that all theoretical works reach the same conclusion makes our
predictions well grounded. $^{176}$Sn is found to be the last
stable nucleus against neutron emission in the Sn isotopic chain
and the predicted halo phenomena in the Zr isotopic chain are
confirmed once again.

Here, it is necessary to mention the mass dependence of the
parameter set TMA. There are several reasons to introduce a mass
dependence to the effective interaction. First, as pointed out in
Ref. \cite{geng.032}, such a mass dependence is indispensable to
reproduce all the nuclei from small to large mass numbers. Second,
since it is well known that the effective interaction is strongly
density dependent, it is reasonable to have a mass dependence to
relate various cluster effects with mass. Third, the good
agreement of many nuclear properties calculated using this
parameter set with those from the parameter set NL3 also supports
such a mass dependence.
\section{Acknowledgments}

L.S. Geng is grateful to the Monkasho fellowship for supporting
his stay at Research Center for Nuclear Physics where this work is
carried out. This work was partly supported by the Major State
Basic Research Development Program Under Contract Number
G2000077407 and the National Natural Science Foundation of China
under Grant No. 10025522, 10221003 and 10047001.

\newpage
\begin{table}[h]
\tbl{The convergence study for $^{132}$Sn with the parameter set
TMA. Listed are the total binding energy, $B$, the binding energy
per nucleon, $B/A$, the charge, neutron, proton, and matter root
mean square radii, $R_c$, $R_n$, $R_p$ and $R_m$, and the
quadrupole deformation parameter for the neutron, proton and
matter distributions, $\beta_{2n}$, $\beta_{2p}$ and $\beta_{2m}$.
$N=N_b=N_f$ is the number of shells used for the expansion of
fermion fields and boson fields.}
{\begin{tabular}{c@{\hspace{1ex}}@{\hspace{1ex}}cc@{\hspace{1ex}}@{\hspace{1ex}}cccc@{\hspace{1ex}}@{\hspace{1ex}}ccc}
\toprule
$N$&$B$&$B/A$&$R_c$&$R_n$&$R_p$&$R_m$&$\beta_{2n}$&$\beta_{2p}$&$\beta_{2m}$\\
\colrule
 20&1103.464&8.360&4.7209&4.9764&4.6526&4.8563&0.000&0.000&0.000\\
 18&1103.544&8.360&4.7203&4.9763&4.6520&4.8560&0.000&0.000&0.000\\
 16&1103.760&8.362&4.7197&4.9750&4.6514&4.8550&0.000&0.000&0.000\\
  14&1103.735&8.362&4.7216&4.9749&4.6533&4.8556&0.000&0.000&0.000\\
   12&1103.258&8.358&4.7211&4.9772&4.6528&4.8569&0.000&0.000&0.000\\
    10&1101.932&8.348&4.7180&4.9669&4.6497&4.8492&0.000&0.000&0.000\\
 \botrule
\end{tabular}}
\end{table}

\begin{table}[htbp]
 \tbl{The same as Table I, but for
$^{176}$Sn.}
{\begin{tabular}{c@{\hspace{1ex}}@{\hspace{1ex}}cc@{\hspace{1ex}}@{\hspace{1ex}}cccc@{\hspace{1ex}}@{\hspace{1ex}}ccc}
\toprule
$N$&$B$&$B/A$&$R_c$&$R_n$&$R_p$&$R_m$&$\beta_{2n}$&$\beta_{2p}$&$\beta_{2m}$\\
\colrule
 20&1174.685&6.674&5.0843&5.7995&5.0209&5.5894&0.000&0.000&0.000\\
 18&1174.621&6.674&5.0841&5.7961&5.0207&5.5868&0.000&0.000&0.000\\
  16&1174.314&6.672&5.0837&5.7886&5.0204&5.5811&0.000&0.000&0.000\\
  14&1174.119&6.671&5.0853&5.7810&5.0220&5.5759&0.000&0.000&0.000\\
    12&1171.623&6.657&5.0862&5.7533&5.0229&5.5556&0.000&0.000&0.000\\
     10&1153.970&6.557&5.0745&5.6764&5.0110&5.4956&0.000&0.000&0.000\\
\botrule
\end{tabular}}
\end{table}

\begin{table}[htbp]
\tbl{The same as Table I, but for $^{130}$Zr.}
{\begin{tabular}{c@{\hspace{1ex}}@{\hspace{1ex}}cc@{\hspace{1ex}}@{\hspace{1ex}}cccc@{\hspace{1ex}}@{\hspace{1ex}}ccc}
\toprule
$N$&$B$&$B/A$&$R_c$&$R_n$&$R_p$&$R_m$&$\beta_{2n}$&$\beta_{2p}$&$\beta_{2m}$\\
\colrule
  20&936.601&7.205&4.6343&5.3251&4.5647&5.1032&-0.218&-0.171&-0.203\\
   18&936.578&7.204&4.6346&5.3171&4.5651&5.0975&-0.215&-0.171&-0.202\\
    16&936.380&7.203&4.6363&5.3036&4.5667&5.0883&-0.213&-0.172&-0.200\\
  14&936.075&7.201&4.6390&5.2893&4.5695&5.0787&-0.208&-0.172&-0.197\\
   12&934.880&7.191&4.6416&5.2637&4.5722&5.0610&-0.199&-0.171&-0.190\\
    10&928.696&7.144&4.6462&5.2038&4.5768&5.0185&-0.181&-0.168&-0.177\\
 \botrule
\end{tabular}}
\end{table}

\begin{figure}[t]
\centering
\includegraphics[scale=0.4]{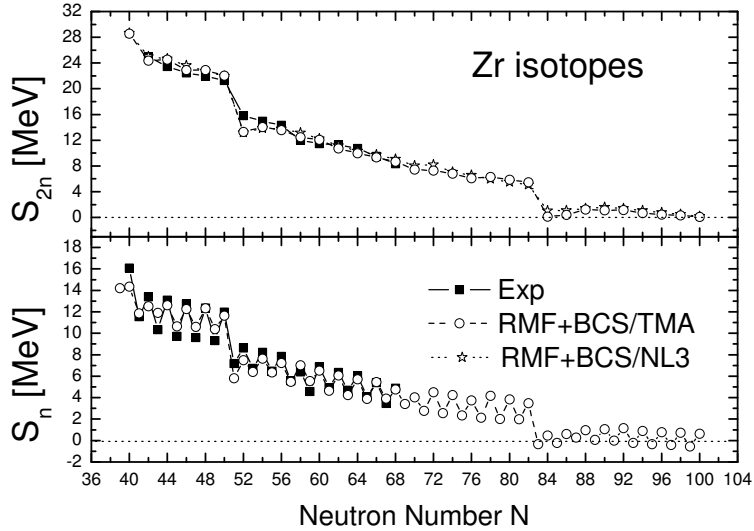}
\caption{The double and single neutron separation energies,
S$_{2n}$ and $S_n$, of Zr isotopes. The results obtained from the
deformed RMF+BCS calculations with TMA set (open circle) are
compared with available experimental data (solid square). In the
upper panel, the results obtained from the deformed RMF+BCS
calculations with NL3 set (open star) are also shown for
comparison.}
\end{figure}

\begin{figure}[h]
\centering
\includegraphics[scale=0.4]{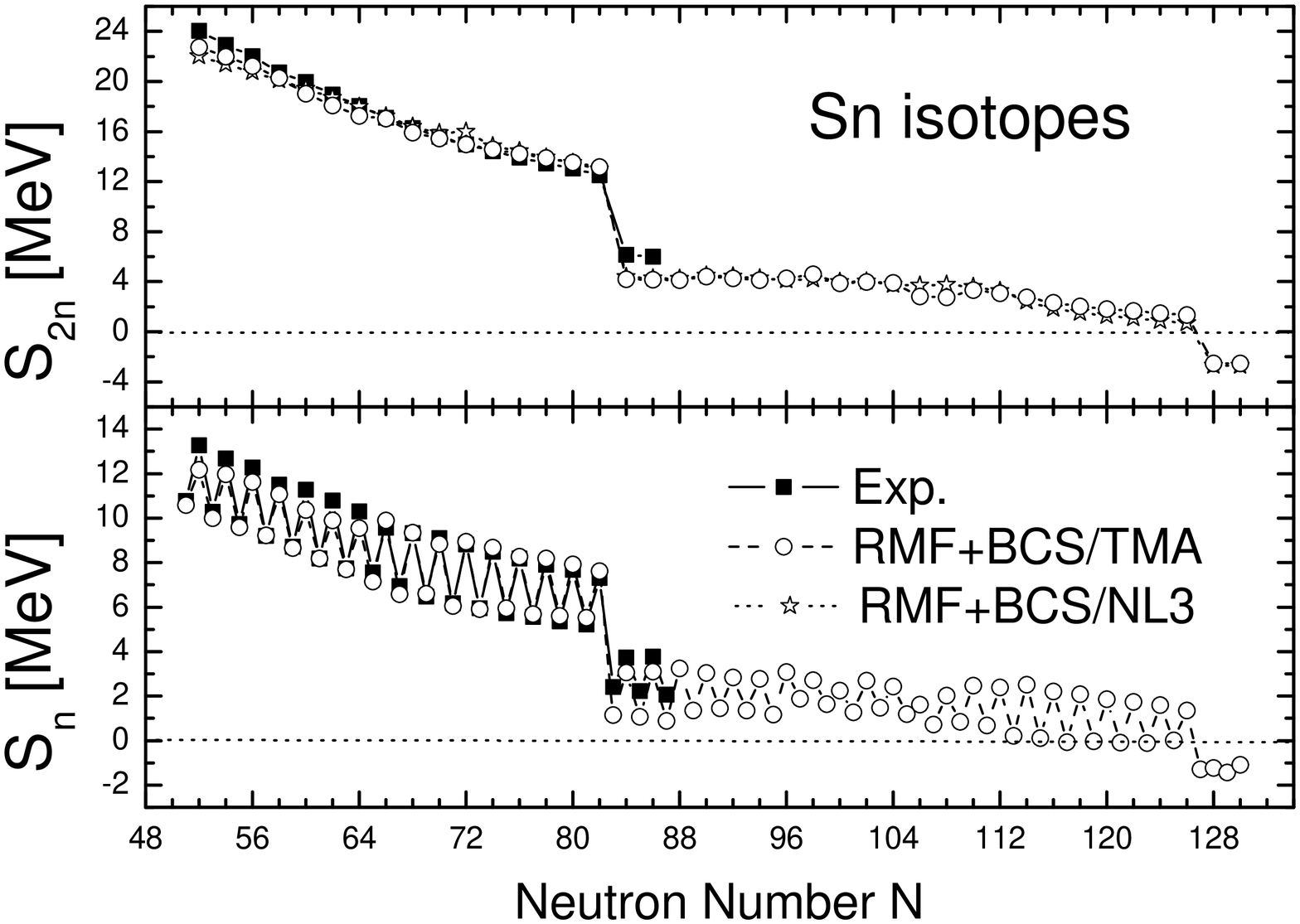}
\caption{The same as Fig. 1, but for Sn isotopes.}
\end{figure}

\begin{figure}[t]
\centering
\includegraphics[scale=0.4]{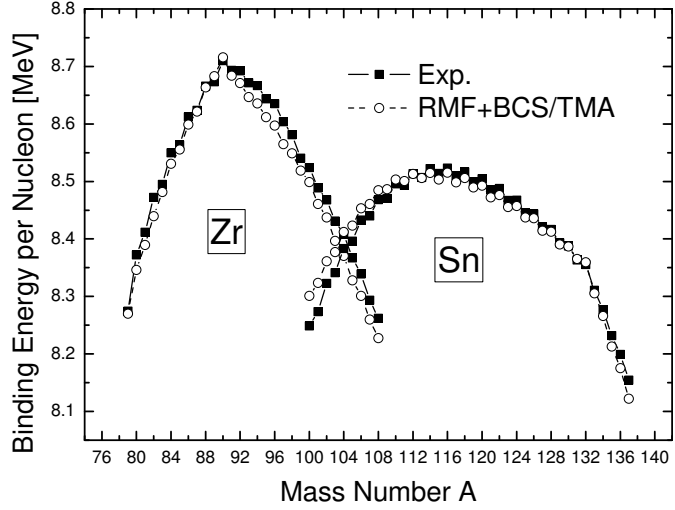}
\caption{\label{fig3.fig} The binding energies per nucleon of Zr
and Sn isotopes. The results obtained from the deformed RMF+BCS
calculations with TMA set (open circle) are compared with
available experimental data (solid square).}
\end{figure}

\begin{figure}[h]
\centering
\includegraphics[scale=0.4]{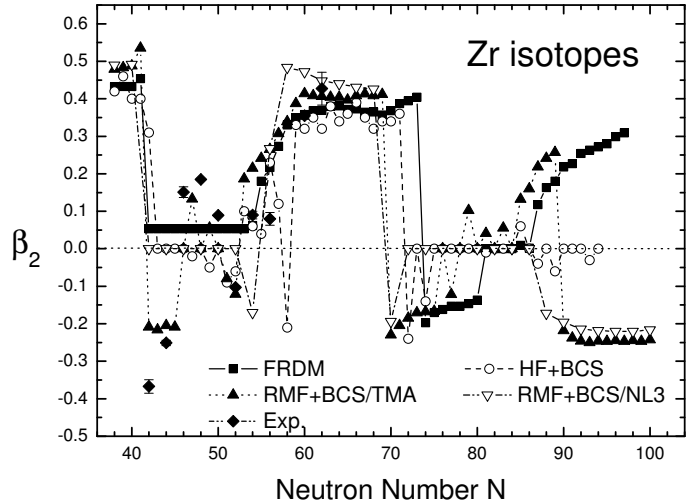}
\caption{\label{fig4.fig}The mass quadrupole deformation
parameters, $\beta_{2}$, of Zr isotopes. The results obtained from
the deformed RMF+BCS calculations with TMA set (solid up-triangle)
and NL3 set (empty down-triangle) are compared with those of
finite range droplet model (FRDM) (solid square), those of Hartree
Fock + BCS (HF+BCS) method (empty circle) and those extracted from
the B(E2) values (solid diamond with error bar). The results from
the RMF+BCS/NL3 calculations are only shown for even-even nuclei.
}
\end{figure}

\begin{figure}[h]
\centering
\includegraphics[scale=0.4]{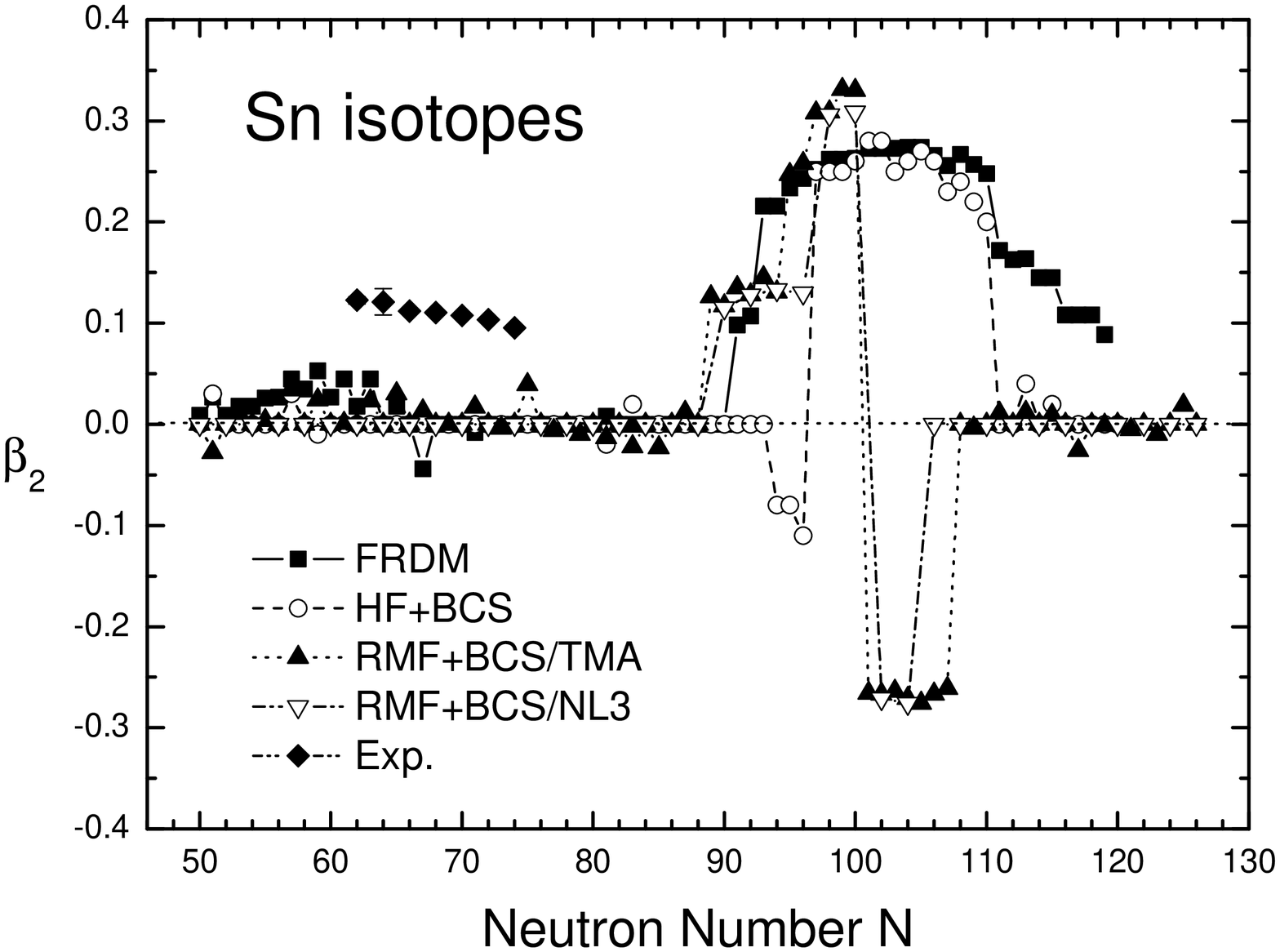}
\caption{The same as Fig. 4, but for Sn isotopes.}
\end{figure}

\begin{figure}[h]
\centering
\includegraphics[scale=0.4]{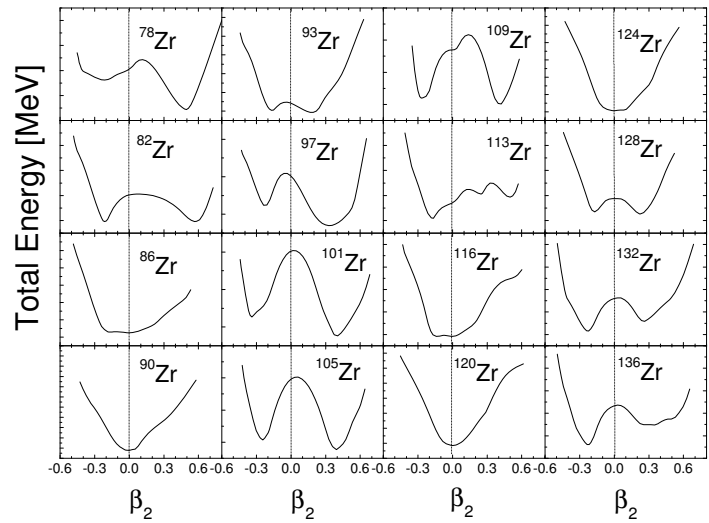}
\caption{\label{fig5.fig}The total energy curves of 16 Zr isotopes
obtained from the deformed RMF+BCS calculations with TMA set as
functions of mass quadrupole deformation parameters, $\beta_{2}$.}
\end{figure}

\begin{figure}[t]
\centering
\includegraphics[height=70mm,width=60mm]{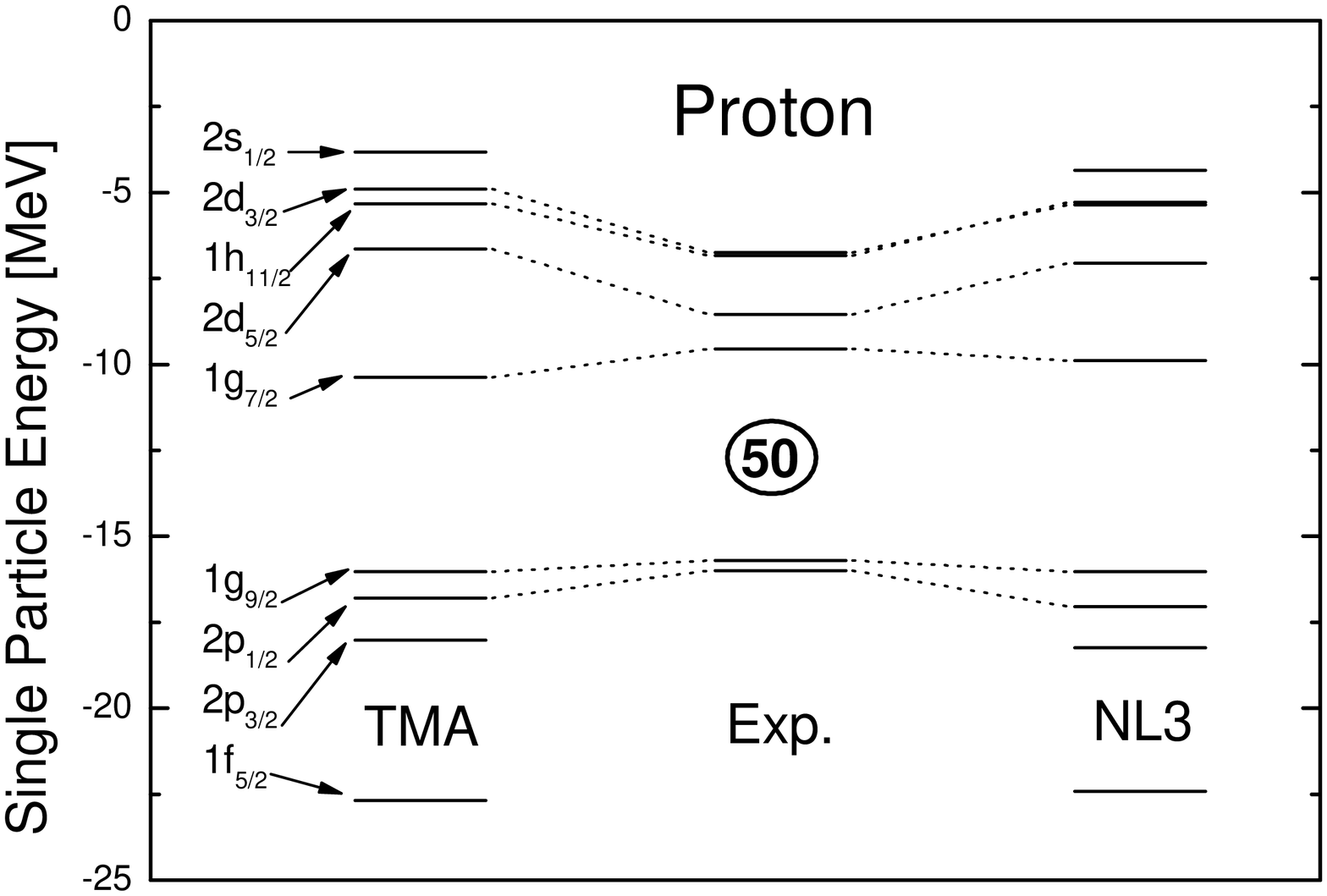}%
\includegraphics[height=70mm,width=60mm]{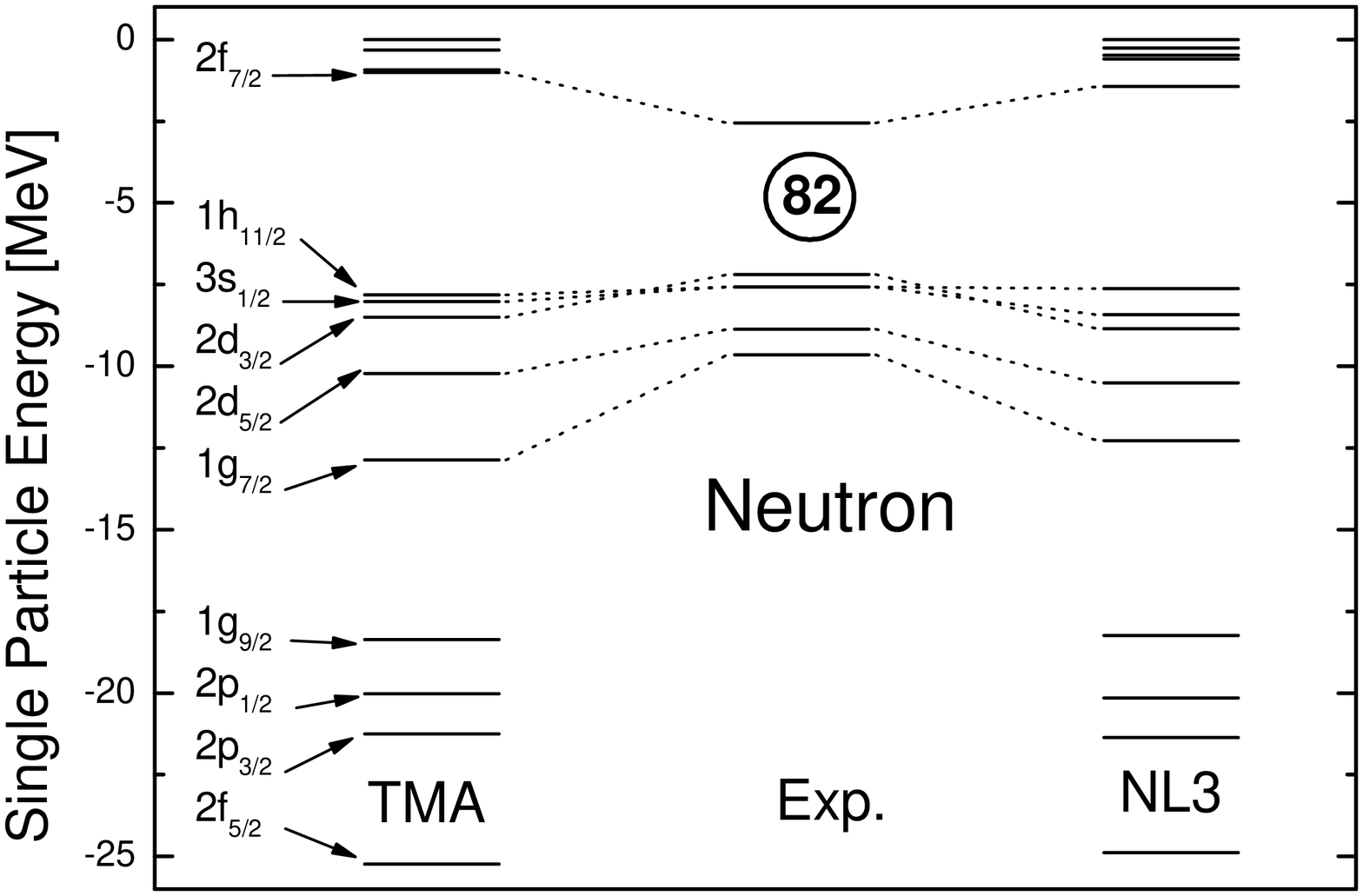}
\caption{\label{fig10.fig}The proton and neutron single-particle
spectra of $^{132}$Sn. The results obtained from the deformed
RMF+BCS calculations with TMA and NL3 parameter sets are compared
with available experimental data.}
\end{figure}

\begin{figure}[t]
\centering
\includegraphics[scale=0.4]{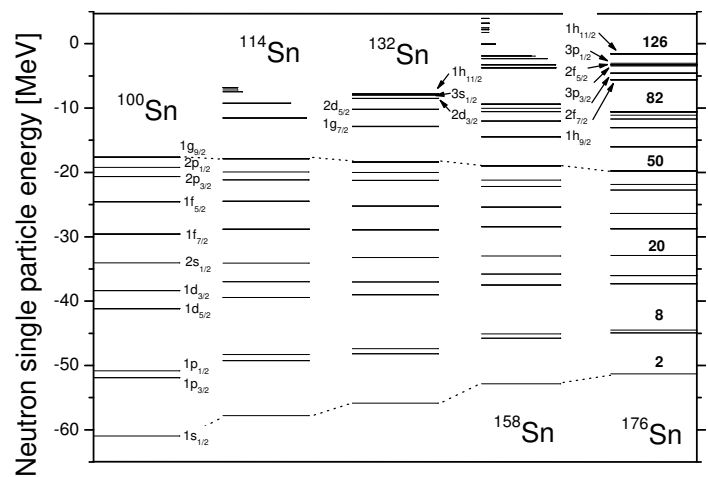}
\caption{\label{fig9.fig}The neutron single-particle spectra of
$^{100}$Sn, $^{114}$Sn, $^{132}$Sn, $^{158}$Sn and $^{176}$Sn
obtained from the deformed RMF+BCS calculations with TMA set.}
\end{figure}

\begin{figure}[h]
\centering
\includegraphics[scale=0.4]{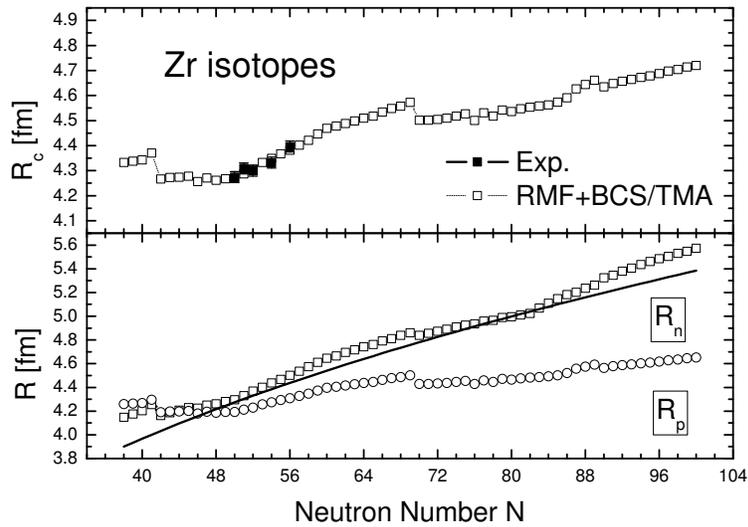}
\caption{\label{fig7.fig}The charge, proton and neutron radii
obtained from the deformed RMF+BCS calculations with the TMA
parameter set (open square and circle) compared with available
experimental data (solid square). The solid line denotes a
$N^{1/3}$ dependence.}
\end{figure}

\begin{figure}[t]
\centering
\includegraphics[scale=0.4]{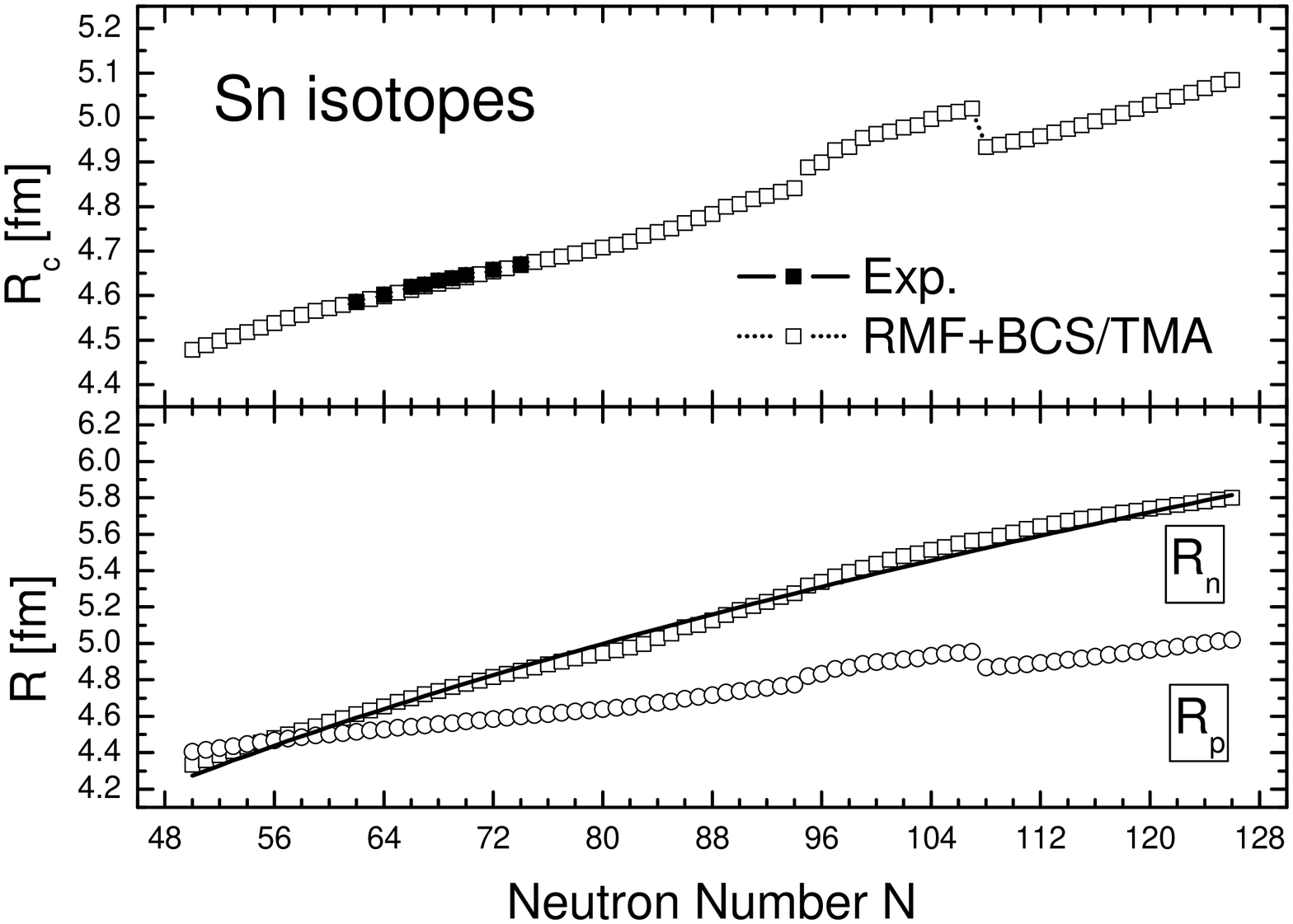}
\caption{The same as Fig. 9, but for Sn isotopes.}
\end{figure}

\end{document}